\documentclass[fleqn,usenatbib]{mnras}
\usepackage{newtxtext,newtxmath}

\usepackage[T1]{fontenc}

\DeclareRobustCommand{\VAN}[3]{#2}
\let\VANthebibliography\thebibliography
\def\thebibliography{\DeclareRobustCommand{\VAN}[3]{##3}\VANthebibliography}

\usepackage{bm}
\usepackage{graphicx}	

\title[The shear modulus of the neutron star crust]{Neutron star crust in Voigt approximation: general symmetry of the stress-strain tensor and an universal estimate for the effective shear modulus}

\author[A.\ I. Chugunov]{
	Andrey I. Chugunov,$^{1}$\thanks{E-mail: andr.astro@mail.ioffe.ru}
	\\
	$^{1}$Ioffe Institute, Politekhnicheskaya 26, 194021 St. Petersburg, Russia
}

\date{Accepted XXX. Received YYY; in original form ZZZ}

\pubyear{2020}

\begin{document}
	\label{firstpage}
	\pagerange{\pageref{firstpage}--\pageref{lastpage}}
	\maketitle

\begin{abstract}

I discuss elastic properties of neutron star crust in the framework of static Coulomb solid model when atomic nuclei are treated as non-vibrating point charges; electron screening is neglected. The results are also applicable for solidified white dwarf cores and other materials, which can be modeled as Coulomb solids (dusty plasma, trapped ions, etc.).  I demonstrate that the Coulomb part of the stress-strain tensor  has additional symmetry: contraction $B_{ijil}=0$. It does not depend on the  structure (crystalline or amorphous) and composition. I show as a result of this symmetry the effective (Voigt averaged) shear modulus of the polycrystalline or amorphous  matter to be equal to $-2/15$ of the Coulomb (Madelung) energy density at undeformed state. This result is general and exact within the model applied. Since the linear mixing rule and the ion sphere model are used, I can suggest a simple universal estimate for the effective shear modulus: $\sum_Z 0.12\, n_Z Z^{5/3}e^2 /a_\mathrm{e}$. Here summation is taken over ion species, $n_Z$ is number density of ions with charge $Ze$. Finally $a_\mathrm{e}=(4 \pi n_\mathrm{e}/3)^{-1/3}$ is electron sphere radius. Quasineutrality condition $n_\mathrm{e}=\sum_Z Z n_Z$ is assumed.
\end{abstract}



\section{Introduction}
A neutron star crust  as well as  a white dwarf core  is made of fully ionized atomic nuclei (ions) and almost unpolarizable background of degenerate electrons (e.g., \citealt{hpy07,ch08,ch17,CompStarBook18}). At  low enough temperature,  which is typical for neutron stars and white dwarfs at their late evolutionary stages,  ions solidify (e.g., \citealt{bst66,oih93,jc96,pc00_EOS_Solid,ch17,mc10_crystallization,Caplan_etal18}). If there is only one type of atomic nuclei at a given layer, they form a crystal with a body-centered cubic (bcc) lattice (\citealt{hfd97,Baiko02,Kozhberov18_Yuk}). However, in  multicomponent models (e.g., \citealt{Fantina_ea20_outer,Carreau_ea20_inner,cfg20}) more complicated lattices are possible (e.g., \citealt{oih93,CF16,Kozhberov19_elast}). Solidification into amorphous structure is discussed elsewhere (e.g., \citealt{Jones99_disord,sca20_disorder}).  

In solidified matter the  shear stresses can be maintained by elasticity, which is important for physics of neutron stars (e.g., \citealt{ch08,ch17}). Namely, elasticity has imprints in oscillation spectra (e.g., \citealt{hc80_torsOsc,st83_tors_osc,McDermott_ea88}; \citealt{Strohmayer_etal91,KY20}) and affects  quasi-periodic oscillations, observed after giant flares of soft gamma repeaters (e.g., 
\citealt{Gabler_ea11,Gabler_ea12,Gabler_ea13,Gabler_ea18}; \citealt{Sotani_ea18}). Elastic stresses can  be responsible for 
asymmetric matter distribution in the crust (so-called mountains; see e.g., \citealt*{Ushomirsky_ea00,hja06_mountains,Horowitz10_lowmass,McDaniel_Owen13}), which is limited by the material strength of the neutron crust matter (see \citealt{hk09,ch10,ch12,bk17,BC18,KY20}). Having this asymmetry, a rapidly rotating neutron star emits gravitational waves;
nowadays gravitational wave observatories put tight  upper bounds for amplitude of such waves, constraining thus the asymmetry
(e.g., \citealt{LIGO_VIRGO20_MSPelipt}). 
Rapid release of magnetic and elastic energy due to crust failure  can be responsible for magnetar bursts
\cite{bl14,bl16,llb16}. Crust failure can be also important for pulsar glitches (e.g., \citealt{bp71_starquake,hm15_glitches,AkbalAlpar18,MelatosDrummond19,ErbilAlpar19,lp20}).

The simplest model to study the elastic properties of the neutron star crust  and  white dwarf core is  Coulomb solid model, which considers a solidified system of point charges (atomic nuclei or ions for brevity) on the uniform neutralizing background (e.g., \citealt{ch17}). Within this model, the elastic properties of a perfect one-component-bcc lattice at zero temperature were calculated by \cite{Fuchs36}. The finite temperature effects were analyzed by \cite{oi90} by Monte Carlo simulations 
and by \cite{Baiko11,Baiko12} within phonon formalism, which naturally includes  quantum effects. Electron screening was considered by \cite{Baiko12,kp15,Baiko15}. Elastic properties of multicomponent crystals were studied by \cite{Kozhberov19_elast}. 

All these papers provide accurate calculations for monocrystals, which 
are not isotropic and whose stress-strain tensor depends on more than two independent  elastic constants. However, the assumption that the whole neutron star crust (or white dwarf core) is one monocrystal seems to be unrealistic. More likely, it should be polycrystalline (e.g., \citealt{kp15,Caplan_etal18}).
Authors also suggest an effective shear modulus $\mu_\mathrm{eff}$, which should be applicable for macroscopically isotropic polycrystalline matter. In early papers, following \cite{oi90}, the effective shear modulus was calculated by averaging the shear wave velocity over directions in crystallite.
As pointed by \cite{kp15}, this approach is equal to so-called Voigt average (respective $\mu_\mathrm{eff}$ is denoted as  $\mu_\mathrm{eff}^\mathrm{V}$ below), which is based on the assumption that  the strain is uniform over all of crystallites;  $\mu_\mathrm{eff}^\mathrm{V}$ known to be an upper limit for $\mu_\mathrm{eff}$ (see \citealt{Blaschke17_elast} and section \ref{Sec:Voigt} for short proof).  \cite{kp15} suggest to apply the self-consistent  theory by \cite{Eshelby61} and calculate respective shear modulus, which turns to be $\sim 22\%$ lower than $\mu_\mathrm{eff}^\mathrm{V}$ for one-component-bcc crystallites.

Unfortunately, as pointed by \citet[section 10, p.\ 36]{ll_elast}, $\mu_\mathrm{eff}$ cannot be calculated accurately, when one has only elastic properties of crystallites, if they are strongly anisotropic (which is the case for one-component Coulomb crystals, e.g., \citealt{Fuchs36,oi90}). The explanation is that $\mu_\mathrm{eff}$ depends on the correlations in orientations of crystallites. The correlations can depend on the evolution history and, as far as I know, currently they cannot be predicted with high degree of reliability. As a result, one should keep in mind an associated uncertainty in $\mu_\mathrm{eff}$, when applying this quantity in astrophysics. 

In this letter, I consider general properties of the Coulomb part of the stress-strain  tensor $B_{ijkl}$ (roman letters $i,\, j,\, k,\ldots$ denote spacial components, see section \ref{Sec:ElastTheory} for details) within Coulomb solid model, neglecting ion vibrations (i.e.\ finite temperature and quantum effects) and assuming the strain to be uniform at the microscopic scale (Voigt appoximation).%
\footnote{
For astrophysical applications the Coulomb solid model represents only (Coulomb) part of the thermodynamics, which is associated with ions (e.g., \citealt{ch08}). To obtain the net pressure, stress-strain tensor, etc.\ one should include contributions associated with degenerate electron gas as well as unbound neutrons in the inner crust of neutron stars. 
These contributions can be considered as additive and isotropic; they do not affect shear modulus (but dominate for pressure and bulk modulus). Thus, Coulomb contribution can be considered separately.
\label{footnote_CoulPart} }
I demonstrate the contraction $B_{ijil}=0$ due to symmetry of the Coulomb interaction (summation over repeated indices is assumed). It is precisely so for the uniform  deformation of arbitrary  (even multicomponent or/and disordered)
The Coulomb solid, which has symmetric stress tensor at undeformed state.
Invariance of this convolution with respect to the Voigt average couples the bulk modulus $K$ and $\mu_\mathrm{eff}^\mathrm{V}$, making only one of them independent. For Coulomb solids, the bulk modulus is determined by the Coulomb (Madelung) energy density of non-deformed state $\epsilon^\mathrm{C}$, so that
$\mu_\mathrm{eff}^\mathrm{V}=-(2/15)\ \epsilon^\mathrm{C}$ (exactly), which can be also treated as an upper limit for $\mu_\mathrm{eff}$.
The linear mixing rule combined with ion sphere model leads to the estimate (\ref{LinMixMu}),  which is applicable for arbitrary composition and structure of the Coulomb solid. 
For neutron star crusts and white dwarf cores, this estimate can be treated as an upper limit for $\mu_\mathrm{eff}$.

\section{Elastisity tensors for Coulomb crystals}
\subsection{Elasticity theory at finite pressure} \label{Sec:ElastTheory}
The elasticity theory describes deformation of solids from some initial (undeformed) state. The deformation can be described by displacement $\bm{\xi}(\bm{R})$ of matter elements  from position $\bm{R}$ to $\tilde{\bm{R}}=\bm{R}+{\bm \xi}(\bm{R})$. For 
infinitesimal deformations 
$\xi(\bm R^a)_i=u_{ij}R^a_j$, where $u_{ij}$ is a displacement gradient. Below I consider a uniform deformation, i.e. $u_{ij}$ assumed to be constant over 
solid (see discussion in section \ref{Sec:Sym}). 

The Coulomb solids have finite pressure in the undeformed state.%
\footnote{
The pressure is formally negative for Coulomb solids (e.g.\ \citealt{ch08}), which does not result in an instability, because the neutralizing background is considered as incompressible. For astrophysical applications, the stability is associated with positive pressure of electron gas (and unbound neutrons in the inner crust of neutron stars).\label{ePresure}}
It makes the elasticity theory lengthier than the standard textbook version (e.g., \citealt{ll_elast}) written for a zero pressure, so that several different tensors should be introduced  (see below for brief summary and \citealt{Wallace67} for details).%
\footnote{\cite*{Marcus_ea02} suggest that the theory of elasticity at finite pressure can be simplified and unified by the use of the Gibbs free energy (see, however, the comment by \citealt{Steinle_Neumann_2004_Marcus_crit} and reply by \citealt{Marcus_Qiu04_reply} for discussion of crucial details). In particular,  the approach by \cite{Marcus_ea02} allows to deal only with the Voigt symmetric tensors. However, at least for some first-principle calculations (e.g.\ \citealt{Baiko11,Baiko12,Kozhberov19_elast} and this work) the approach by \cite{Wallace67} seems to be more useful because it shortens derivations.}
The first one,  $S_{ijkl}$ describes the change of the energy $\delta E$ (per unit volume $V$ of undeformed matter), associated with the deformation:
\begin{equation}
\delta E=\sigma_{ij} u_{ij}+\frac{1}{2} S_{ijkl} u_{ij} u_{kl}. \label{S_def}
\end{equation}
Here $\sigma_{ij}$ is the stress tensor in an undeformed solid, assumed to be isotropic below 
($\sigma=-P\delta_{ij}$, where $P$ is pressure, $\delta_{ij}$ is Kronecker delta).
Generally, $S_{ijkl}$ does not have Voigt symmetry (e.g., \citealt{Wallace67}).

The tensor $S_{ijkl}$ should be distinguished from the stress-strain tensor $B_{ijkl}=S_{ijkl}-
P\left(\delta_{il}\delta_{jk}
-\delta_{ij}\delta_{lk}\right)$, which is used to calculate the change of the stress tensor $\delta \sigma_{ij}$, associated with the deformation
\begin{equation}
\delta \sigma_{ij}=\frac{1}{2} B_{ijkl}\left(u_{kl}+u_{lk}\right).
\label{B_def}
\end{equation}
Straightforward calculations demonstrate that 
%
$B_{ijkl}+B_{ilkj}=S_{ijkl}+S_{ilkj}$. 
The tensor $B_{ijkl}$ has Voigt symmetry
($B_{ijkl}=B_{jikl}=B_{ijlk}=B_{klij}$, see \citealt{Wallace67}), and, thus, up to 21 independent elastic parameters.

For isotropic material, the stress-strain tensor $B^\mathrm{V}_{ijkl}$ has the same structure as for the elasticity theory at a zero pressure:
\begin{equation}
B^\mathrm{V}_{ijkl}=K\delta_{ij}\delta_{kl}
+\mu^\mathrm{V}_\mathrm{eff}\left(\delta_{ik}\delta_{jl}+\delta_{il}\delta_{jk}
-\frac{2}{3}\delta_{ij}\delta_{kl}\right),
\label{BV}
\end{equation}
giving a common form of the stress-strain relation
\begin{eqnarray}
\delta \sigma_{ij}
=K\delta_{ij} u_{ll}
+\mu^\mathrm{V}_\mathrm{eff}\left(u_{ik}+u_{ki}-\frac{2}{3}\delta_{ik}u_{ll}\right).
\end{eqnarray}

\subsection{Symmetry of the elasticity tensor for Coulomb crystals} 
\label{Sec:Sym} 
To derive $S_{ijkl}$ tensor, I calculate a change of energy associated with deformation. The energy of the Coulomb solid can be written as
\begin{eqnarray}
E&=&\sum_{a} \sum_{b>a}
\frac{Z^a Z^b e^2}{\left|\bm R^{a}-\bm R^{b}\right|}
-  \sum_a\int \frac{Z^a e^2 n_\mathrm{e} }{\left|\bm R^{a}-\bm r\right|}\mathrm d^3 \bm r
\nonumber \\
&+&\frac{1}{2}
\int\int  \frac{e^2 n_\mathrm{e}^2}{\left|\bm r- \bm r^\prime\right|} 
\mathrm d^3 \bm r\,
\mathrm d^3 \bm r^\prime.
\label{energy}
\end{eqnarray}
Here upper indices $a$ and $b$ enumerate ions, $Z^a e$ and $\bm R^a$ are the position and the charge of ion $a$ respectively ($e$ is an absolute value of electron charge).
The electron number density is uniform; $n_\mathrm{e}(\bm r)=\sum_a Z^a/V$ due to quasineutrality condition ($V$ is volume). 
In equation (\ref{energy}) terms are ion-ion, ion-electron, and electron-electron interaction energies respectively.
The ion positions after the deformation are
\begin{equation}
\tilde{R}_i^a=R_i^a+u_{ij}R^a_j.
\label{tildeRa}
\end{equation}
The same deformation is applied for electrons. As a result, their number density becomes $\tilde n_e =n_e/J$, where $J$ is Jacobian of $\bm r \rightarrow \tilde {\bm r}$ transformation (the quasineutrality condition obviously holds true after deformation).
The energy in the deformed state is
\begin{eqnarray}
\tilde E&=&\sum_{a} \sum_{b>a}
\frac{Z^a Z^b e^2}{\left|\tilde {\bm R}^{a}-\tilde {\bm R}^{b}\right|}
-  \sum_a\int \frac{Z_i e^2 \tilde n_\mathrm{e}}{\left|\tilde {\bm R}^{a}-\tilde {\bm r}\right|}\mathrm d^3 \tilde {\bm r}
\nonumber \\
&+&\frac{ 1}{2}
\int\int \frac{e^2 \tilde n_\mathrm{e}^2}{\left|\tilde {\bm r}- \tilde {\bm r^\prime}\right|} 
\mathrm d^3 \tilde {\bm r}\,
\mathrm d^3 \tilde {\bm r^\prime}.
\label{tenergy}
\end{eqnarray}
Formal change of variables in all integrals from $\tilde {\bm r}$ to $\bm r$ leads to appearance of a Jakobian factor $J$, which finally disappears, because $J\tilde n_e =n_e$.
Thus, the change of the energy associated with the deformation can be written as
\begin{eqnarray}
\delta E&=& \tilde E-E=
\sum_{a} \sum_{b>a}
Z^a Z^b e^2\left(\frac{1}{|\tilde \Delta |}-\frac{1}{| \Delta |}\right)
\nonumber \\
&-& 
\sum_a\int Z_i e^2 n_\mathrm{e} \left(\frac{1}{|\tilde \Delta |}-\frac{1}{| \Delta |}
\right)\mathrm d^3 \bm r
\nonumber \\
&+&\frac{1}{2}
\int\int  e^2 n_\mathrm{e}^2
\left(\frac{1}{|\tilde \Delta |} -\frac{1}{| \Delta |} \right)
\mathrm d^3 \bm r\,
\mathrm d^3 \bm r^\prime.
\label{denergy}
\end{eqnarray}
Here and below in the first line  $\bm \Delta=\bm R^{a}-\bm R^{b}$, in the  second line  $\bm \Delta=\bm R^{a}-\bm r$, and,  in the third line $\bm \Delta=\bm r-\bm r^\prime$. Similar notations are applied for quantities in the deformed state, e.g.,  $\tilde {\bm \Delta}=\tilde{\bm R}^{a}-\tilde{\bm R}^{b}$ for the first line.

The Taylor expansion over $\bm {\xi}$ for terms in parenthesis in the equation (\ref{denergy}) gives:
\begin{equation}
\frac{1}{|\tilde { \Delta}|}-\frac{1}{| \Delta |}\approx
	\frac{\Delta_i \Delta_j}{\Delta^3} u_{ij}
	+\frac{3 \Delta_i \Delta_k-\Delta^2 \delta_{ik}}
	{\Delta^5}\Delta_j \Delta_l u_{ij}
	u_{kl}.
	\label{Taylor}
\end{equation}
Comparison  with the equation (\ref{S_def}) gives:
\begin{eqnarray}
\sigma_{ij}&=& 
\sum_{a} \sum_{b>a}
Z^a Z^b e^2 \frac{\Delta_i \Delta_j}
{\Delta^3}
- 
\sum_a\int Z_i e^2 n_\mathrm{e}  \frac{\Delta_i \Delta_j}
{\Delta^3}  \mathrm d^3 \bm r
 \nonumber \\
&+&\frac{1}{2}
\int\int  e^2 n_\mathrm{e}^2
 \frac{\Delta_i \Delta_j}
{\Delta^3} \mathrm d^3 \bm r\,
\mathrm d^3 \bm r^\prime.
 \label{sigma_ij}
 \\
S_{ijkl}&=& 
\sum_{a} \sum_{b>a}
Z^a Z^b e^2 \frac{3 \Delta_i \Delta_k-\Delta^2 \delta_{ik}}
{\Delta^5}\Delta_j \Delta_l 
\nonumber \\
&-& 
\sum_a\int Z_i e^2 n_\mathrm{e} \frac{3 \Delta_i \Delta_k-\Delta^2 \delta_{ik}}
{\Delta^5}\Delta_j \Delta_l  \mathrm d^3 \bm r
\label{S_ijkl} \\
&+&\frac{1}{2}
\int\int   e^2 n_\mathrm{e}^2
\frac{3 \Delta_i \Delta_k-\Delta^2 \delta_{ik}}
{\Delta^5}\Delta_j \Delta_l \mathrm d^3 \bm r\,
\mathrm d^3 \bm r^\prime.
\nonumber
\end{eqnarray}
To calculate $S_{ijil}=S_{ijkl}\delta_{ik}$ , I
perform the contraction before the summation or/and integration: each term contains $(3 \Delta_i \Delta_k-\Delta^2 \delta_{ik})\delta_{ik}\equiv 0$, thus $S_{ijil}=0$.%
\footnote{This result can not be  straightforwardly  generalized for general pow-law potential $\propto 1/r^n$, because for this potential the respective term becomes $\left[n(n+2) \Delta_i \Delta_k-n \Delta^2 \delta_{ik}\right]\delta_{ik}= n(n-1)\Delta^2\ne 0$ for $n\ne 1$.}
In particular, $S_{ijij}=0$ (for cubic symmetry this form is equivalent to the previous).
It is easy to show that  $B_{ijil}=0$ (in particular, $B_{ijij}=0$).

In this section I do not appeal to any assumption on a structure or a composition, thus the results can be applied for multicomponent Coulomb solids  with arbitrary structure  (crystalline or disordered).

It is worth to warn the reader that in this section I assume that the deformation is uniform at the \textit{microphysical} level, i.e.\ displacement of all ions is given by equation (\ref{tildeRa}). This assumption seems natural, especially when considering deformation of crystals (e.g., \citealt{Baiko15}). However, generally it can be violated -- for given macroscopic  deformation (e.g., applied at the boundary) the deformation field within the solid can be non-uniform, if it is energeticaly favorable (i.e.\ if it leads to a change of the energy lower than given by the equation \ref{tenergy}). In this case, the actual $S_{ijkl}$ tensor corresponds to lower energy and relations  $S_{ijil}=0$ and $B_{ijil}=0$ can be violated.

\subsection{Shear modulus for isotropic material: the Voigt average }
\label{Sec:Voigt}
Here I apply the well known Voigt average  approach (e.g.\ \citealt{Blaschke17_elast}) to estimate  the elastic properties of the polycrystalline matter.
The approach is based on the assumption that the strains are equal for all crystallites, leading to
the second-order change in energy in the form
\begin{equation}
 \delta^{(2)} E=\sum_c \frac{V_c}{2\,V} S^{c}_{ijkl} u_{ij}u_{kl}
 =S^\mathrm{V}_{ijkl} u_{ij}u_{kl},
\end{equation}
where summation is performed over crystallites, $V_c$ and  $S^{c}_{ijkl}=S_{mnop} 
R^c_{im}R^c_{jn}R^c_{ko}R^c_{lp}$ are volume and elastic tensors for the crystallite $c$. 
Here rotation matrix $R^c_{im}$ is applied to transform the original frame to the crystal frame.
$S^\mathrm{V}_{ijkl}$ corresponds to the Voigt averaged elastic tensor.
The Voigt average gives an upper limit for $\mu_\mathrm{eff}$ due to the same reasons as  discussed at the end of the previous section: it assumes the uniform deformation for all crystallites, but non-uniform deformation, in principle, can decrease the energy and, subsequently, lead to a lower actual value of $\mu_\mathrm{eff}$. 
Two contractions are invariant with respect to Voigt average:
$S_{iijj}^\mathrm{V}=S_{iijj}$ and $S_{ijij}^\mathrm{V}=S_{ijij}$
 (e.g., \citealt{Blaschke17_elast}; the invariance follows from $R_{ik}R_{il}=\delta_{kl}$, see e.g.,  \citealt*{vms88}).
As far as the second one vanishes for Coulomb solids, it should also vanish for the Voigt-averaged stress-strain tensor: $S_{ijij}^\mathrm{V}=0$. 

It is worth to point, that  $S_{ijij}^\mathrm{V}=0$ can be derived  directly within the approach of the section \ref{Sec:Sym} considering deformation of polycrystalline matter as a whole.

Contraction $S_{ijij}^\mathrm{V}=0$ implies  $B_{ijij}^\mathrm{V}=0$.
Combined with general form of $B_{ijkl}$ for isotropic material (equation \ref{BV}) it imposes condition
\begin{equation}
 \mu^\mathrm{V}_\mathrm{eff}=-\frac{3}{10}K=-\frac{2}{15}E^\mathrm{M},
 \label{result_mu}
\end{equation}
which should hold true exactly.
In the last equality I use $K=4 \epsilon^\mathrm{M}/9$, which follows from scaling of the Madelung energy density
$\epsilon^\mathrm{M}\propto
n_\mathrm{e}^{4/3}$. 
As far as $\epsilon^\mathrm{M}$ is negative (see footnote \ref{ePresure}),  $\mu^\mathrm{V}_\mathrm{eff}$ is positive.

Let me check the equation (\ref{result_mu}) by previous calculations.
To begin with, the Madelung energy  for
one-component-bcc crystal
was calculated by \cite{Baiko_ea01}:
$\epsilon^\mathrm{M}=-0.895929255682\,Z^{5/3} e^2 n_\mathrm{Z} /a_\mathrm{e}$, where $a_\mathrm{e}=\left[3/(4\pi n_\mathrm{e})\right]^{1/3}$ is electron sphere radius
and $n_\mathrm{Z}$ is number density of ions with charge $Ze$.  
According to equation (\ref{result_mu}):
\begin{equation}
\mu^\mathrm{V}_\mathrm{eff,\ bcc}
=0.119457234091\frac{Z^{5/3} e^2}{a_\mathrm{e}}n_Z. \label{mu_bcc}
\end{equation}
It perfectly agrees with the result by \cite{Baiko11}, where the coefficient $0.1194572$ was given.
For two-component ordered crystals, tensor $S_{ijkl}$ was calculated by \cite{Kozhberov19_elast} and $S_{ijij}=0$ 
(within this calculation's accuracy).%
\footnote{In fact, some miraculous coincidences in the numerical results  by \cite{Kozhberov19_elast} were the starting point for this study.}  

Similarly, the equation (\ref{result_mu}) can be straightforwardly applied to calculate $\mu^\mathrm{V}_\mathrm{eff}$  for multicomponent lattice structures, whose energies were calculated by \cite{CF16,Kozhberov18_Yuk}, but $\mu^\mathrm{V}_\mathrm{eff}$ was not studied previously.

Even more important is the fact that the equation (\ref{result_mu})  allows me
to write down an universal estimate for $\mu^\mathrm{V}_\mathrm{eff}$ using the ion-sphere model by \cite{Salpeter54} combined with linear mixing rule (e.g., \citealt{HV76,oih93}). This combination 
describes Madelung energy of crystals very accurately
(\citealt{oih93,Kozhberov19_elast}) and leads to 
\begin{equation}
\epsilon^\mathrm{M}\approx -\frac{9}{10}\sum_Z  \frac{Z^{5/3} e^2}{a_\mathrm{e}}\, n_Z.
\label{LinMixMad}
\end{equation}
According to the equation (\ref{result_mu}):
\begin{equation}
\mu^\mathrm{V}_\mathrm{eff}\approx -\frac{3}{25}\sum_Z  \frac{Z^{5/3} e^2 }{a_\mathrm{e}}\,n_Z
=0.12\sum_Z  \frac{Z^{5/3} e^2}{a_\mathrm{e}}\, n_Z.
\label{LinMixMu}
\end{equation}
For one-component-bcc it deviates from the accurate calculations by (\ref{mu_bcc}) for mere $0.05\%$.

The results of this section (equations \ref{result_mu} and \ref{LinMixMu}) are applicable for amorphous Coulomb solids, which are generally isotropic and do not require Voigt average.

\section{Summary and discussion} \label{Sec:Concl}
I consider elastic properties of neutron star crust matter within the Coulomb solid model (solidified system of point charges (ions) on the uniform unpolarizable neutralizing background).
The results are also applicable for white dwarf cores and other systems, which can be described by Coulomb solid model (e.g., dusty plasma). 

Neglecting thermal and zero-point vibrations of ions
and assuming uniform strain, 
I derive universal exact relation for the Coulomb part of the stress-stain tensor $B_{ijil}=0$ (also for tensor in equation \ref{S_def}: $S_{ijil}=0$). 
This result
does not depend on the structure and composition of the solid.

I apply this result to consider elastic properties of macroscopically isotropic polycrystalline matter using Voigt average
and demonstrate that the effective shear modulus is determined by the Madelung energy density $\epsilon^\mathrm{M}$: $\mu^\mathrm{V}_\mathrm{eff}=-(2/15)\,\epsilon^\mathrm{M}$ (again, exactly and for arbitrary structure and composition).
I check this relation by comparison with previously calculated values for one- and two-component crystals.
Using the ion sphere model by \cite{Salpeter54} and linear mixing rule to estimate Madelung energy, I suggest a simple estimate for $\mu^\mathrm{V}_\mathrm{eff}$ (equation \ref{LinMixMu}). The results are also directly applicable for isotropic amorphous  solids.

The Voigt average is equal to averaging of the dispersion relations for long-wavelength transversal modes (approach suggested by \citealt{oi90}) and thus $\mu^\mathrm{V}_\mathrm{eff}$ can be applied to calculate root-mean-square velocity of these modes in monocrystals.
For amorphous solids, which are isotropic,  equation (\ref{LinMixMu}) allows to consider long-wavelength transversal modes and thus discuss their  low-temperature thermodynamics.

Universal estimate (\ref{LinMixMu}) suggests that the elastic properties should vary rather smoothly  within neutron star crust, at least if there are no rapid changes of composition. This result is important for torsional oscillations of neutron star crust (e.g.\ \citealt{KY20}). Equation   (\ref{LinMixMu}) also supports all previous neutron star models based on the effective shear modulus of one-component-bcc crystal: their numerical results are valid for arbitrary microscopic  structure of the crust, because  $\mu^\mathrm{V}_\mathrm{eff}$ depends on it weakly.

It is worth to warn the reader, that all results in this letter were obtained assuming uniform deformation at the microphysical level (i.e., displacement of all ions is given by equation \ref{tildeRa}). As discussed at the end of section \ref{Sec:Sym}, this assumptions can be violated, if it  allows the lower energy at the same (macroscopic) strain. In this case $S_{ijil}$ can be non-zero. 
In particular, for polycrystalline and amorphous matter the estimate (\ref{LinMixMu}) should be considered as an upper limit for the effective shear modulus (the Voigt average known to give an upper limit for the effective shear modulus).
However, as follows from \cite{Kozhberov19_elast}, multicomponent Coulomb crystals tend to be less anisotropic than one-component crystals and different estimates of $\mu_\mathrm{eff}$ become closer to $\mu_\mathrm{eff}^V$.
The ion vibrations and electron screening are neglected in this work. They decrease the shear modulus (e.g., \citealt{Baiko12}), thus equation (\ref{LinMixMu}) gives an upper bound for $\mu_\mathrm{eff}$ even if these effects are included.

The approach of the section \ref{Sec:Sym} can be easily generalized for the non-spherical nuclei (so-called pasta phases) in the bottom of the inner crust of neutron star by introducing proton charge density with respective integration instead of summation over point-like ions.
I plan to study the applicability of such model for pasta phases in subsequent publication.

Similarly, the result of the section  \ref{Sec:Sym} can be applied in electrostatics: the variation of electrostatic energy of a body with a given charge distribution under uniform deformation can be described by an analogue of the equation (\ref{S_def}), where contraction  $S_{ijil}=0$.

\section*{Acknowledgements}
I'm grateful to A.A.~Kozhberov, who provided me with an unpublished (at that time) version of  \cite{Kozhberov19_elast}. I also thank anonymous
referee for the fast and constructive report.
This research was partially supported by The Ministry of Science and
Higher Education of the Russian Federation (Agreement with Joint
Institute for High Temperatures RAS No 075-15-2020-785).

\section*{DATA AVAILABILITY}
The data underlying this letter are available in the letter.

\bibliographystyle{mnras}

\bsp	
\label{lastpage}
\end{document}